\documentclass[prl, preprintnumbers, twocolumn, amsmath, amssymb, superscriptaddress]{revtex4}
\usepackage{graphicx}
\usepackage{dcolumn}
\usepackage{bm}

\begin{document}

\title{Robust Signatures of Majorana Fermions in a Semiconductor Quantum Ring}
\author{Aram Manaselyan}
\affiliation{Department of Solid State Physics, Yerevan State
University, Yerevan, Armenia}
\author{Areg Ghazaryan}
\affiliation{Department of Physics and Astronomy, University of
Manitoba, Winnipeg, Canada R3T 2N2}
\author{Tapash Chakraborty\footnote{Tapash.Chakraborty@umanitoba.ca}}
\affiliation{Department of Physics and Astronomy, University of
Manitoba, Winnipeg, Canada R3T 2N2}

\begin{abstract}
We have investigated the possible presence of Majorana fermions in a semiconductor quantum ring 
containing a few interacting electrons, and a strong spin-orbit interaction, proximity coupled 
to an $s$-wave superconductor. We have found that for rings with sizes of few hundred angstroms 
and for certain values of the chemical potential and the entire range of the magnetic field, there 
are strong indications of the presence of Majorana fermions. In particular, the ground state 
energies and the average electron numbers for the states with even and odd electron numbers are 
almost identical. We have also studied the wave functions of Majorana fermions in the 
ring and have shown that Majorana fermions are well separated from each other in the angular coordinates. 
As the semiconductor quantum rings with a few interacting electrons are available in the laboratories, 
we believe that the long sought-after Majorana fermions could perhaps be unequivocally observed in 
such a system.

\end{abstract}

\maketitle

Search for Majorana fermions (MFs), the particles that are their own antiparticles and are governed 
by non-Abelian exchange statistics \cite{Majorana,wilczek} has intensified in recent years. The hybrid 
semiconductor-superconductor nanostructured systems \cite{Oreg,Lutchyn,Alicea,Flensberg} 
are believed to be the most likely systems hosting such exotic fermions \cite{note}, and have quite 
naturally received considerable attention by various experimental groups \cite{Mourik,Deng1,Das,Deng2,Churchill}. 
Experimental efforts have also focused on the ferromagnetic atomic chain \cite{Nadj-Perdge} which is 
in close proximity to a conventional superconductor. A promising route for realization of the MFs 
\cite{Oreg,Lutchyn} is the observation of the topological superconducting phase in a one-dimensional 
semiconductor wire  with large Rashba spin-orbit (SO) coupling \cite{Bychkov}, proximity coupled to 
an $s$-wave superconductor. By tuning the chemical potential of the system in the gap region created 
by an applied magnetic field, the system is effectively rendered spinless and the MFs are expected 
to reside at the edges of the wire, akin to Kitaev's original $p$-wave superconductor 
chain model \cite{KitaevModel}. In addition to discovering the telltale signs of this
long sought-after particle, as it was pointed out earlier by various authors, one 
major impetus for discovering the MFs lies in their potential use in topological quantum 
computation \cite{NayakReview,SternReview,AliceaNetwork} because of their unusual exchange
statistics. Here we show that, instead of a quantum wire, semiconductor quantum rings with 
their doubly connected geometry and consequent unique quantum properties reveal a much 
stronger signature of the presence of MFs. Observation of the Aharonov-Bohm oscillations 
\cite{Aharonov} and the persistent current \cite{Buttiker} in small semiconductor quantum 
rings (QR), and recent experimental realization of QRs with only a few electrons \cite{Lorke} 
have made QRs an attractive topic of experimental research and a unique playground for various 
many-body effects in these quasi-one-dimensional systems \cite{Chakraborty}. We demonstrate below 
that a semiconductor QR containing a few electrons, proximity coupled to an $s$-wave superconductor, 
could be an excellent candidate for detecting signatures of Majorana fermions since the energy 
spectrum of such system has a lot of level crossing due to Aharonov-Bohm oscillations. This periodic
energy spectrum for a few interacting electrons entails the suitable conditions required, in particular, 
with the help of the magnetic field we can bring two energy levels with even and odd parity close
to each other thereby facilitating the existence of MF in a QR.

In what follows we consider a two-dimensional QR with internal radius $R^{}_1$ and external 
radius $R^{}_2$ with strong Rashba SO coupling \cite{Bychkov}, proximity coupled to an $s$-wave 
superconductor. We chose the confinement potential of the QR to be infinitely high borders: 
$V^{}_{\rm conf}(\rho)=0$, if $R^{}_1\leq \rho\leq R^{}_2$ and infinity outside of the QR. Without 
the superconducting pairing potential the Hamiltonian of the system is, ${\cal H}=\sum_{i}^{N^{}_e}{\cal 
H}_\mathrm{SP}^i+\frac12\sum_{i\neq j}^{N^{}_e}V^{}_{ij}$. Here $N^{}_e$ is the number of electrons in 
the QR, $V^{}_{ij}=e^2e^{-\lambda r}/\epsilon \left|\mathbf{r}^{}_i-\mathbf{r}^{}_j\right|$ is the Yukawa 
type screened Coulomb interaction term with screening parameter $\lambda$ \cite{majorana_prb} and 
${\cal H}^{}_\mathrm{SP}$ is the single-particle Hamiltonian in the presence of an external perpendicular 
magnetic field and with the SOI included,
\begin{equation}\label{singleH}
{\cal H}_{\rm SP}={\frac1{2m^{}_e}}\Pi_i^2-\mu+V_{\rm
conf}(\rho)+\frac12 g\mu^{}_B B\sigma^{}_z+H^{}_{\rm SO},
\end{equation}
where $\bf \Pi=\bf p-\frac ec \bf A$, $\bf A$ is the vector potential of the magnetic field along the 
$z$ axis, and 
$\mu$ is the chemical potential. The third term on the right hand side of (\ref{singleH}) is the
Zeeman splitting. The last term describes the Rashba SOI \cite{Bychkov}
\begin{equation}\label{Rashba}
H^{}_{\rm SO}=\frac \alpha\hbar \left[{\boldsymbol
\sigma}\times\left({\bf p}-\frac ec {\bf A}\right)\right]^{}_z,
\end{equation}
with $\alpha$ being the SOI parameter, which is sample dependent and is proportional to the interface 
electric field that confines the electrons in the $xy$ plane. In Eq.~(\ref{singleH}) and Eq.~(\ref{Rashba}),
$\boldsymbol \sigma$ is the electron spin operator and $\sigma^{}_x,\sigma^{}_y$ and $\sigma^{}_z$ are 
the Pauli spin matrices. We take as the basis states the eigenstates of ${\cal H}_{\rm SP}$ for $B=0$ 
and $\alpha=0$ \cite{AregPhysE}. In doing so we can cast the many-body hamiltonian ${\cal H}$ into the second
quantized form and diagonalize it to get the eigenstates and the eigenvectors \cite{majorana_prb}.

The proximity induced superconductivity potential can then be written directly in this basis 
${\cal H}^{}_\mathrm{SC}=\Delta\sum^{}_n\left(c^{}_{n\downarrow}c^{}_{n \uparrow}+c^\dagger_{n\uparrow} 
c^\dagger_{n\downarrow}\right),$ where for brevity we write $n=\{n_e,l_e\}$, where $n^{}_e$ and $l^{}_e$ 
are electron radial and angular quantum numbers. The pairing potential strength $\Delta$ is taken to be real. 
In order to evaluate the eigensates of the total hamiltonian ${\cal H}^{}_\mathrm{PSC}={\cal H}+{\cal 
H}^{}_\mathrm{SC}$ we again use the exact diagonalization procedure to diagonalize ${\cal H}^{}_\mathrm{PSC}$ 
in even and odd sectors \cite{majorana_prb}. For example, for the odd sector we diagonalize ${\cal 
H}^{}_\mathrm{PSC}$ for a system with non-constant number of electrons, namely $1,3, \dots N^{}_e$ electron 
number basis. A similar procedure is employed for the even sector as well. This gives us the possibility to 
obtain the low-lying energy states and the wave functions both for even and odd sector very accurately.
A major advantage of using the exact diagonalization scheme in a QR over that in a quantum wire is that the
convergence is much better here without the requirement of an induced gap in the spectrum \cite{majorana_prb}. 

\begin{figure}
\includegraphics[width=8cm]{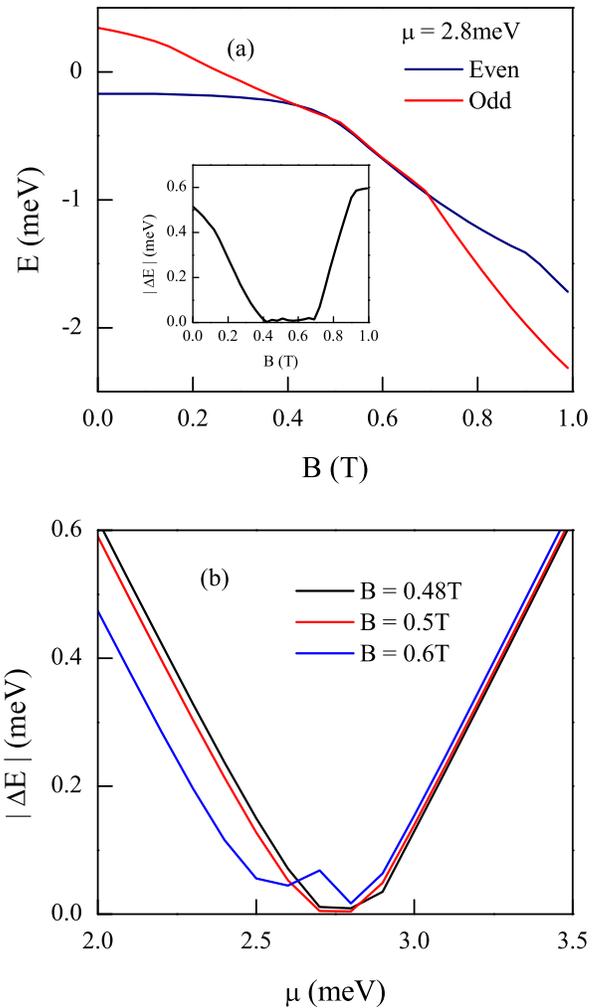}
\caption{\label{fig:EdepBMu} (a) The dependence of the energies of the ground states of odd and even sector
on the magnetic field $B$. Inset: absolute difference between the energies of the ground states. (b) The
dependence of the absolute difference between the energies of the ground states of odd and even sector on
the chemical potential $\mu$ for various values of the magnetic field.}\end{figure}

We employ several different approaches to identify the signatures of topological superconductivity and the 
existence of well separated MFs in the system. In condensed matter physics, isolated MFs are zero-energy 
quasiparticle excitations and they do not carry a charge \cite{Alicea}. Therefore adding a non-local electron 
which is comprised of two well separated MFs will not alter both the total energy and the charge of the system. 
Even in the case of the system without boundaries, where isolated MFs usually reside, the phase transition 
between the trivial and non-trivial (topological) superconducting states results in closing of the superconducting
bulk gap. Based on this premise the first parameter which is used for identifying the phase transition between 
two superconducting phases and the appearance of isolated MFs is the energy difference between the odd and even 
sector \cite{Stoudenmire} 
\begin{equation}
\Delta E=\left|E^{}_\mathrm{odd}-E^{}_\mathrm{even}\right|. 
\end{equation}
This quantity is expected to vanish in the topological phase, but remains finite for the ordinary superconducting 
state \cite{Stoudenmire}. The next parameter is the charge difference between even and odd sector $\Delta N$, 
which is equal to the mean electron number difference between the two sectors. In order to calculate this 
parameter we first calculate the particle densities in each sector 
\begin{equation}
\rho^{}_\mathrm{even,odd}(\mathbf r)=\int d\mathbf{r}^{}_2d\mathbf{r}^{}_3\dots
d\mathbf{r}^{}_{N^{}_e}\left|\Psi\left(\mathbf{r},\mathbf{r}^{}_2,\dots,\mathbf{r}^{}_{N^{}_e}
\right)\right|^2,
\end{equation}
where
$\Psi\left(\mathbf{r}^{}_1,\mathbf{r}^{}_2,\dots,\mathbf{r}^{}_{N^{}_e}\right)$
is the wave function of the system in odd and even sector based on the parity of $N^{}_e$. It is 
known that for a semiconductor quantum wire in the topological superconducting phase, changing the parameters
of the system, such as the chemical potential or the magnetic field strength, results in a change of the ground 
state parity. For a finite size wire this is accompanied by a jump of the total electron number and the 
charge due to the jump is spread along the wire and has an oscillating behavior \cite{Ben-Shach}. Therefore 
we first calculate the difference between the particle densities in odd and even sector $\Delta\rho(\mathbf{r})=
\rho^{}_\mathrm{odd}(\mathbf{r})-\rho^{}_\mathrm{even}(\mathbf{r})$ and compare our results for the ring with 
those of a quantum wire. The charge difference between odd and even sector $\Delta N$ is the cumulative difference 
between particle densities, i.e., $\Delta N=\int d\mathbf{r}\Delta\rho(\mathbf{r})$. Finally, we can also
calculate the MF probability distributions directly, using the procedure outlined previously 
\cite{majorana_prb,Stoudenmire}
\begin{equation}\label{MWF}
p^{(a)}(\rho,\varphi)=\sum^{}_s\left|\sum^{}_{n}d^{(a)}_{ns}
\phi^{}_n\left(\rho,\varphi\right)\right|^2,
\end{equation}
where $d^{(a)}_{ns}$ are the expansion coefficients of the linear expansion of the MFs operators $\gamma^{}_a$ 
in terms of the electron creation and annihilation operators $c^\dagger_{ns}$ and $c^{}_{ns}$. Here $s$ denotes 
the spin quantum number of the electron and $a=1$ and $a=2$ corresponds to left and right Majorana
edge states respectively.

\begin{figure}
\includegraphics[width=8cm]{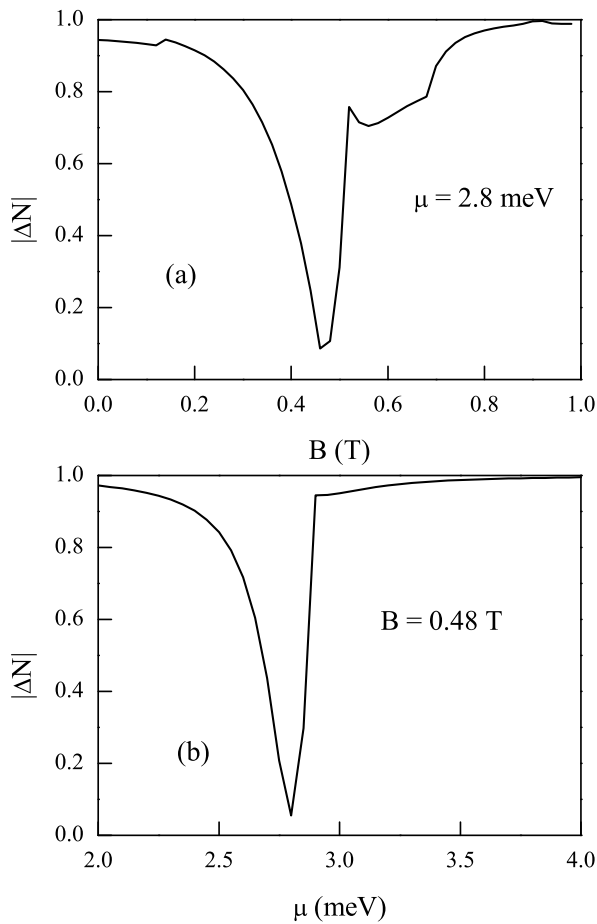}
\caption{\label{fig:DNdepBMu} (a) The absolute difference between the average electron numbers of the ground
states of odd and even sector as a function of (a) the magnetic field $B$ and (b) the chemical potential.
}
\end{figure}

For our present work we consider the InAs semiconductor QR with parameters: $m=0.042m^{}_0$, where $m^{}_0$ 
is the bare electron mass, $g=-14$, $\epsilon=14.6$ \cite{QDspinorbit}, and the SO coupling strength
$\alpha=20\,\mathrm{meV\cdot nm}$ \cite{Fasth}. We take the superconducting pairing potential strength to be 
$\Delta=0.225$ meV. In our calculations, $R^{}_1=30$ nm and $R^{}_2=80$ nm, and we have considered the 
screening parameter to be $\lambda=0.1\,\,\mathrm{nm^{-1}}$.
In Fig. 1 (a), the magnetic field dependence of the ground state energies for the even and odd sectors are 
presented for the chemical potential $\mu=2.8$ meV. The figure clearly illustrates that there is a range of the 
magnetic field from 0.4 T to 0.7 T where the energies of even and odd sectors are very close to each other and
have very similar behavior. The absolute value of the difference of these energies against the magnetic field is 
presented as an inset. In the range $B=0.4-0.7$ T the energy difference has the oscillatory behavior. In Fig.
1 (b), the absolute value of the difference of even and odd sectors ground state energies are presented against 
the chemical potential for three different values of the magnetic field. Again it is clear from the figure that 
there is a range of the chemical potential from 2.5 to 3 meV where the energy difference is very small and
also has an oscillatory behavior. It should also be noted that starting from the value of the chemical potential 
$\mu=2.8$ meV the dependence of the energy difference against $\mu$ is almost independent of the magnetic field. 
All of these facts are the first signatures of the presence of Majorana fermions in our semiconductor QR.

In order to confirm that we indeed have signatures of Majorana fermions in our QR we have presented in Fig.~2 
the absolute difference of the average electron numbers of even and odd sector ground states versus (a) the
magnetic field and (b) the chemical potential. Figure 2 clearly shows that for $\mu=2.8$ meV and $B=0.48$ T 
the average electron numbers of even and odd sector ground states are approximately equal to each other. This 
is another strong indication for the presence of Majorana fermions in our QR. The probability distribution of 
the Majorana fermion inside the QR is presented in Fig.~3 for $\mu=2.8$ meV and $B=0.48$ T, obtained using 
Eq.~(\ref{MWF}). As it can be seen from the figure, two Majorana peaks are highly separated from each other
by the angular coordinate $\varphi$, but both of them are in the central part of the ring. For the first
peak $\varphi=\pi/2$ while for the second peak $\varphi=3\pi/2$. Interestingly, the very presence of a central 
barrier of the ring (not present in a quantum wire or a quantum dot \cite{QD}) prevents overlap of two MFs while
being so close.

\begin{figure}
\includegraphics[width=8cm]{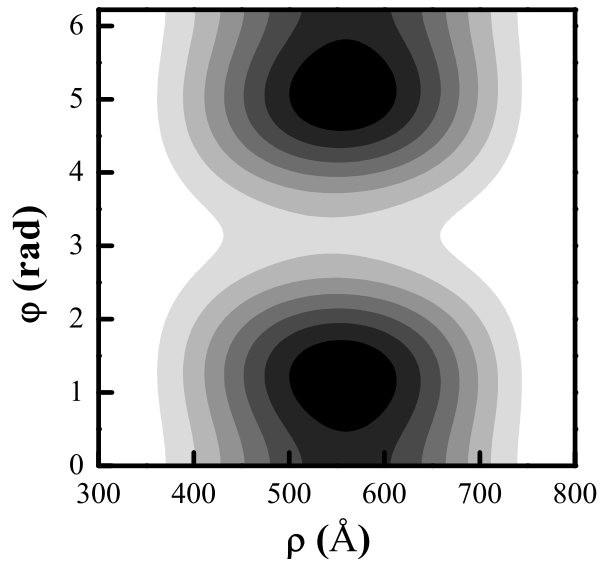}
\caption{\label{fig:Majorana} Majorana fermion probability distribution inside
the ring for $B=0.48$ T and $\mu=2.8$ meV.}
\end{figure}

\begin{figure}
\includegraphics[width=8cm]{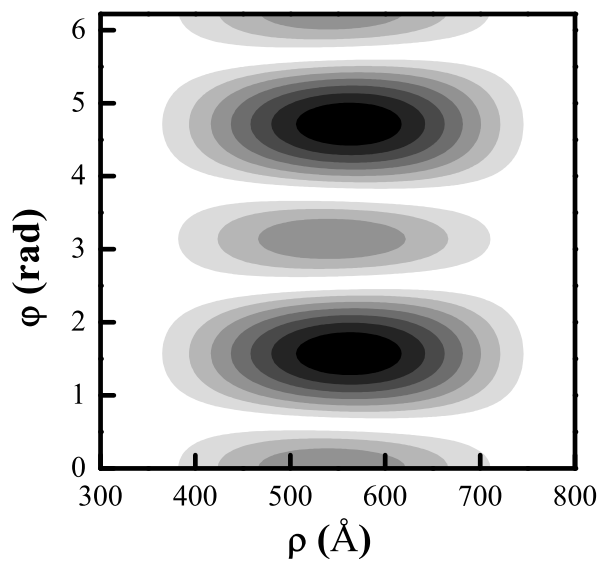}
\caption{\label{fig:DensContour} The difference between the single-particle densities of the many-body
states
in odd and even sector for $B=0.48$ T and $\mu=2.8$ meV.}
\end{figure}

Finally, in Fig.~\ref{fig:DensContour} the contour plot of the difference between the single-particle densities 
of the many-body states in odd and even sectors is presented for the parameter values $B=0.48$ T and $\mu=2.8$ 
meV. As we see in Fig.~\ref{fig:DNdepBMu}, for these parameters the mean electron number difference in odd and
even sectors is very small, but is not equal to zero. The contour plot in Fig.~\ref{fig:DensContour} shows how 
this charge difference is distributed in the ring. In fact, this figure indicates that it is mostly localized 
in the center part of the ring in the $\rho$ direction but is spread through whole ring region in the $\varphi$ 
direction and has the oscillatory behavior. This result is in accord with the result for a semiconductor wire  
\cite{Ben-Shach}, which once more confirms that the obtained results are in fact signatures of the emergence of
topological superconductivity and isolated MFs. In Fig.~\ref{fig:DensContour} the screened Coulomb interaction 
is included in the calculation. We have done similar calculations for the non-interacting case and have seen 
that interaction does not play a major role in the observed distribution, which is also the case in a quantum 
wire \cite{Ben-Shach}.

In conclusion, we have studied here the electronic states in a few-electron semiconductor quantum ring with 
a strong SOI and proximity coupled to an $s$-wave superconductor. We have shown that there are very strong 
signatures of the presence of Majorana fermions in such a system. In particular, we have shown that for the 
entire ranges of the magnetic field and for certain values of the chemical potential, the differences of the 
ground state energies and the average electron numbers for odd and even sectors is close to zero and have 
the oscillatory behavior. Further, we have shown that for certain values of the chemical potential and the 
magnetic field, two Majorana fermions are largely separated from each other in the angular coordinate, but 
both of them are located in the center of the ring. We believe that in many ways, few-electron quantum rings 
are more appropriate for locating the MFs than the quantum wires or quantum dots. Optical spectroscopy or 
magnetotransport measurements on quantum rings \cite{Lorke} can provide important information on the energy 
spectrum. There were some theoretical studies of Majorana fermions and topological phase transitions in 
`superconducting' rings \cite{Pientka,Rosenstein,Lee,Scharf}, but finding appropriate materials in that 
case would be a major challenge. On the other hand, semiconductor quantum rings containing a few electrons 
are in fact, available in the laboratories, and we believe that signatures of Majorana fermions could be 
finally observed in such a system.

The work has been supported by the Canada Research Chairs Program of the Government of Canada
and by the Armenian State Committee of Science Research grant (Project No. 13YR- 1C0014).

\end{document}